\documentclass[%
 reprint,
 amsmath,amssymb,
 aps,
 prb,
]{revtex4-2}

\usepackage{graphicx}
\usepackage{dcolumn}
\usepackage{bm}
\usepackage{hyperref}
\usepackage{xcolor}
\usepackage{hyphenat}
\usepackage[squaren]{SIunits}
\usepackage{subfig}
\usepackage{array,cases}
\usepackage{paralist}
\usepackage{ulem}

\begin{document}

\preprint{APS/123-QED}

\title{Early Stage Growth of Amorphous Thin Film:\\Average Kinetics, Nanoscale Dynamics and Pressure Dependence}

\author{Chenyu Wang}
\email{cywang96@bu.edu}
\affiliation{Department of Physics, Boston University, Boston, MA 02215}

\author{Karl F. Ludwig, Jr.}
\email{ludwig@bu.edu}
\affiliation{Department of Physics and Division of Materials Science and Engineering, Boston University, Boston, MA 02215}

\author{Christa Wagenbach}
\altaffiliation{Current Address: Thermo Fisher Scientific Inc.}
\author{Meliha G. Rainville}
\affiliation{Division of Materials Science and Engineering, Boston University, Boston, MA 02215}

\author{Suresh Narayanan}
\author{Hua Zhou}
\affiliation{Advanced Photon Source, Argonne National Lab, Lemont, IL 60439}

\author{Jeffrey G. Ulbrandt}
\author{Randall L. Headrick}
\affiliation{Department of Physics and Materials Science Program, University of Vermont, Burlington, VT 05405}%

\par
\date{\today}

\begin{abstract}
We used the Coherent Grazing Incidence Small Angle X-Ray Scattering (Co-GISAXS) technique to study the average kinetics and nanoscale dynamics during early-stage a-WSi\(_2\) sputter deposition. The kinetic and dynamic properties are examined as a function of pressure, which is known to be a critical factor in determining final surface roughness. Surface growth kinetics and dynamics are characterized by time parameters extracted from the height-height structure factor and correlation functions. The roughness at a given length scale reaches a maximum before relaxing down to a steady state. The lateral length scale dependence and pressure dependence are then compared among measured kinetics and dynamics time parameters. Surfaces grown at lower pressures are smoother, leading to longer correlation times. The time to reach a dynamic steady state and a kinetic steady state show contrasting pressure dependence. A dynamic steady state is reached earlier than the kinetic steady state at high pressure. A more random deposition direction and lower kinetic energy at higher pressures can explain these phenomena, along with the hypothesis that larger nanoclusters form in vapor before arriving at the surface. A continuum model is applied to simulate the overall behavior with mixed success.
\end{abstract}

\maketitle

\section{\label{sec:Intro}Introduction}
The high level of materials science development in recent years enables the ``materials by design" concept \cite{olson2000designing}. In-depth understanding of the structure and surface growth is a prerequisite to design materials with novel functions to meet the increasing industrial and societal needs. Non-epitaxial thin film growth exhibits qualitatively universal morphologies which have long been encoded in structure zone models based upon empirical relationships between growth conditions and post-facto observations of film structure \cite{movchan1969fiz,thornton1977high}. During a thin film growth process, the final structure arises from the kinetics of the average surface morphology evolution, which is in turn determined by local dynamical processes. In this context, the average \textit{kinetics} means the spatially averaged evolution over the nanoscale surface structure, and the local \textit{dynamics} means the nanoscale fluctuation in temporal evolution \cite{myint2021nanoscale}. At the atomic scale these dynamical processes include adatom arrival, diffusion, incorporation, and desorption \cite{barabasi1995fractal}. During ongoing growth of the thin film, we distinguish here between obtaining a kinetic steady state, when there is no longer evolution of the average structure at a given length scale, and obtaining the dynamic steady state, when the spectrum of local dynamic processes reaches a steady state.\par

When used in a surface sensitive mode, X-ray Photon Correlation Spectroscopy (XPCS) enables the analysis of dynamical process in the near surface structure. Coherent X-ray scattering can produce a speckle pattern, which reflects the temporal configuration in the coherent illuminated volume \cite{headrick2019coherent}. Under coherent illumination, scattering from surface height fluctuations forms a speckle pattern on the detector, and one can infer the dynamical processes leading to such evolution \cite{sutton2008review}. In this paper, we use XPCS to examine the relationship between average surface structure and the underlying dynamics during early stage sputter deposition of a-WSi$_2$ as a function of growth pressure. The technique of Grazing Incidence Small Angle X-Ray Scattering (GISAXS) is applied to improve the surface sensitivity. It is known that growth pressure plays a critical role during the growth of a-WSi$_2$, with a transition occurring between relatively smooth growth surfaces at low pressure and rough growth surfaces at higher pressure \cite{zhou2010pressure}.\par

A number of theories of amorphous thin film growth have been developed to explain the morphology evolution \cite{kuramoto2003chemical,sivashinsky1983instabilities,raible2000amorphous_minimal,raible2000amorphous_theory,raible2001amorphous,pelliccione2008evolution,villain1991continuum,lai1991kinetic,mokhtarzadeh2017simulations,sneppen1992dynamic}. Nonlinear continuum models are the most concise method to simulate growth surface evolution. The data obtained in these XPCS experiments provide considerable information about evolving kinetic and dynamic properties of the growing film in \(k\)-space. Therefore a model of surface growth was simulated and the results were compared with the X-ray scattering. With the implementation of further analysis, we can then discuss the ability of the continuum model to reproduce the experimental results. \par

The rest of the paper is organized as follows. In Section \ref{sec:Exp}, we introduce the main experimental technique we used, Co-GISAXS, and give a detailed description of the sputter deposition experiments under different pressure conditions. In Section \ref{sec:Anal}, the data taken under three pressure conditions are analyzed in both kinetics and dynamics aspects. The height-height structure factor, two-time correlation function and \(g_2\) autocorrelation function are calculated to extract different time parameters which characterize the early stage behavior. In Section \ref{sec:Sim}, we discuss a continuum model of amorphous thin film growth and the simulation results. The experimental and simulation results are compared in this section. In Section \ref{sec:Conc} comprehensive conclusions are given. 

\section{\label{sec:Exp}Experiments}

\subsection{\label{subsec:Co-GISAXS}XPCS and GISAXS}

Photon Correlation Spectroscopy (PCS) provides information on the dynamic behavior of condensed matter across a wide range of length- and time-scales. Since the wavelength of hard X-rays lies in the subnanometer range, XPCS can provide such information at the nanoscale. XPCS measures spontaneous density fluctuations of a material. In the kinetic steady state, conventional X-ray scattering fails to provide such information about a material growth since incoherent X-ray beam measures only the ensemble-averaged density correlation function \cite{sandy2018hard}.\par

Fig.\ref{fig:geom} is a schematic diagram of Co-GISAXS geometry. $\mathbf{k}_i$ and $\mathbf{k}_f$ are the wave vectors of incident and scattered beams, making angles $\alpha_i$ and $\alpha_f$ relative to the surface. The change of wave vector during the scattering is
\begin{equation}
    \mathbf{q} = \mathbf{k}_f - \mathbf{k}_i
    \label{eq:q}
\end{equation}
and the magnitude of \(\mathbf{q}\) is denoted as \(q\). In order to obtain the scattering pattern with surface information, the use of a grazing incidence angle is required. During the diffuse scattering process, there are X-rays propagating along the surface. The beam exits at an angle of total external reflection, forming the band of enhanced scattering in Fig.\ref{fig:geom}, known as the Yoneda wing \cite{yoneda1963anomalous}. When \(\alpha_i\leqslant\alpha_c\) and \(\alpha_f\leqslant\alpha_c\), where \(\alpha_c\) is the critical angle of total reflection, the scattering depth is at the order of a few nanometers. Under such circumstances, the scattered beam is surface sensitive \cite{renaud2009probing}, so it is not mixed with much bulk information during the growth. Co-GISAXS is a non-destructive technique which can be performed as a real-time measurement.

\subsection{\label{subsec:expdet}Experimental details}
\begin{figure}
    \centering
    \includegraphics[width=3.375in]{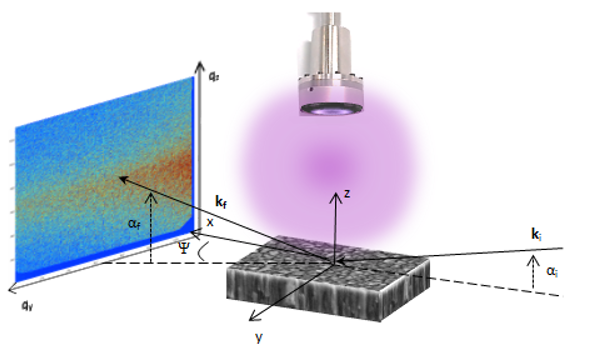}
    \caption{The experimental setup for GISAXS studies during WSi$_2$ sputter deposition}
    \label{fig:geom}
\end{figure}

Coherent X-ray experiments were performed in a GISAXS geometry at the 8-ID-I beamline of \nohyphens{the Advanced Photon Source} (APS) (Fig.\ref{fig:geom}) at \nohyphens{Argonne National Laboratory}. The scattering in the horizontal direction on the detector is primarily in the plane of the sample, and for simplicity the wavenumber of scattering in this direction will be referred to as $q_{\parallel}$. The energy of the incoming X-rays was $7.38~\kilo\electronvolt$ and the focused beam was $20~\micro\metre$ horizontal and $4~\micro\metre$ vertical. The X-ray flux at the sample was $\sim7\times10^{10}$ photons/\second. Scattering was captured by a two-dimensional Princeton Instruments LCX-1300 CCD detector, located $4067~\milli\metre$ from the sample. The detector has 1340 pixels horizontal and 1300 pixels vertical with each pixel being $20\times20~\micro\metre^2$. In the experimental geometry, it covered a range of $q_{\parallel}$ of $0.23\sim0.47~\nano\metre^{-1}$. Data collection scans were 1024 frames in length, with each frame having 1 second of data and followed by a 1 second readout time of the detector. The X-ray incidence angle, $\alpha_i$, was varied around \(0.45\degree\), always remaining close to and above the critical angle $\alpha_c = 0.4\degree$ of the sample. The relationship between $\alpha_i$ and $\alpha_c$ could ensure scattering from the surface instead of the bulk of the film. For the data analyzed here, the X-ray exit angle, $\alpha_f$, was centered around the Yoneda peak.\par

The base pressure of the custom UHV chamber was $7\times10^{-7}~\mathrm{Torr}$ and it used $2^{\prime\prime}$ MeiVac MAK DC Sputter source at a working distance of $10~\centi\metre$. The sputter deposition source is located at a position normal to the sample surface. The target was pre-sputtered for at least $30~\minute$ prior to deposition with the sample shutter closed to remove contamination from the target. Deposition was performed at room temperature onto Si substrates with a $500~\nano\metre$ thermal SiO$_2$ layer. Pressures of 16 mTorr (Sample 1), 10 mTorr (Sample 2) and 8 mTorr (Sample 3) were used with ultra-high purity argon (99.999\%) as the working gas. It is known that there is a transition around 5 mTorr growth pressure between smooth films below this value to films which exhibit increasing roughness with increasing pressure \cite{salditt1996observation}. However, due to the decreased X-ray scattering from smoother films, growth at 8 mTorr was the lowest for which we could obtain sufficient signal for XPCS experiments.\par

For a given deposition power, the deposition rate was approximately independent of pressure. Sample 2 and 3, grown at $25~\watt$ of deposition power, had a growth rate of $0.18~\nano\metre/\second$. Sample 1 was grown at $50~\watt$. Doubling the deposition power increases the deposition rate by a factor of two and likewise decreases the growth time constants by a factor of two. Real-time GISAXS measurements show that surface evolution stops when deposition ends, suggesting that the dominant surface evolution at this temperature is driven by the deposition itself and that thermal effects at this temperature are minimal. We therefore do not expect the difference in deposition rates to affect the fundamental evolution of the surface structure, but only its rate. To aid the reader in comparability of the samples, the timescales from Sample 1 will be scaled by a factor of two so that samples are directly comparable using ``effective time''. The ``effective time'' is used as the default time scale except in Fig.\ref{fig:I-t}.

\section{\label{sec:Anal}Data Analysis}

\subsection{\label{subsec:I}Height-height structure factor}

In the experimental geometry (Fig.\ref{fig:geom}), the projected direction of the incident beam on the sample surface is taken to be the \(x\)-axis, and the \(z\)-axis points normally out of the surface. The \(y\)-axis is taken to be perpendicular to the \(x\) direction in the sample surface. Therefore, the Yoneda wing extends along the \(y\) direction in Fig.\ref{fig:geom} and the detector is in the \(yz\)-plane. When surface features are not too high, the intensity \(I(\mathbf{q},t)\) on the Yoneda wing is approximately proportional to the square modulus of the Fourier transformation of the surface height \cite{rainville2015co}, i.e., \(I(\mathbf{q},t) \propto S(\mathbf{q},t)\), where the height-height structure factor \(S(\mathbf{q},t)\) is defined as
\begin{equation}
    S(\mathbf{q},t) = h(\mathbf{q},t)h^*(\mathbf{q},t)
    \label{eq:hhcorr}
\end{equation}
where \( h(\mathbf{q},t)\) is the Fourier transformation of \( h(x,y,t)\).\par

\begin{figure}
    \centering
    \subfloat[]{\includegraphics[width=1.685 in]{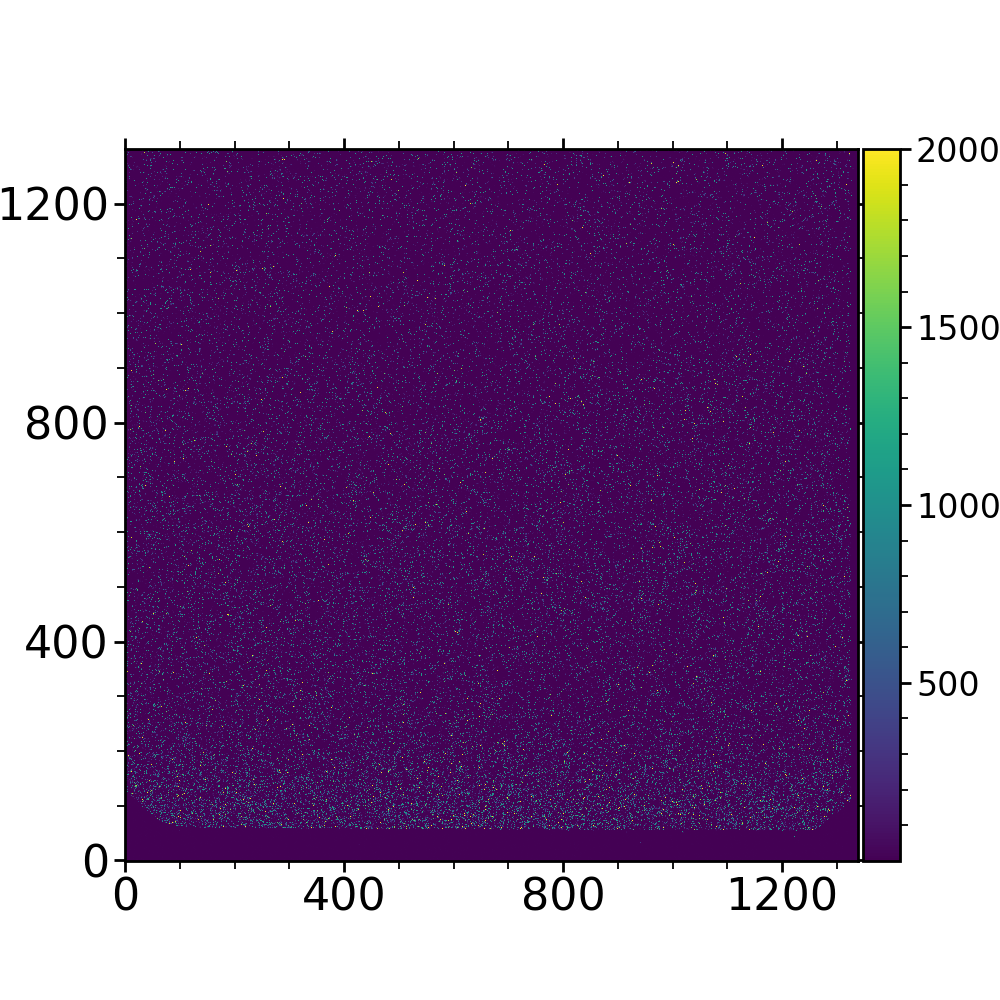}\label{subfig:fr121}}
    \subfloat[]{\includegraphics[width=1.685 in]{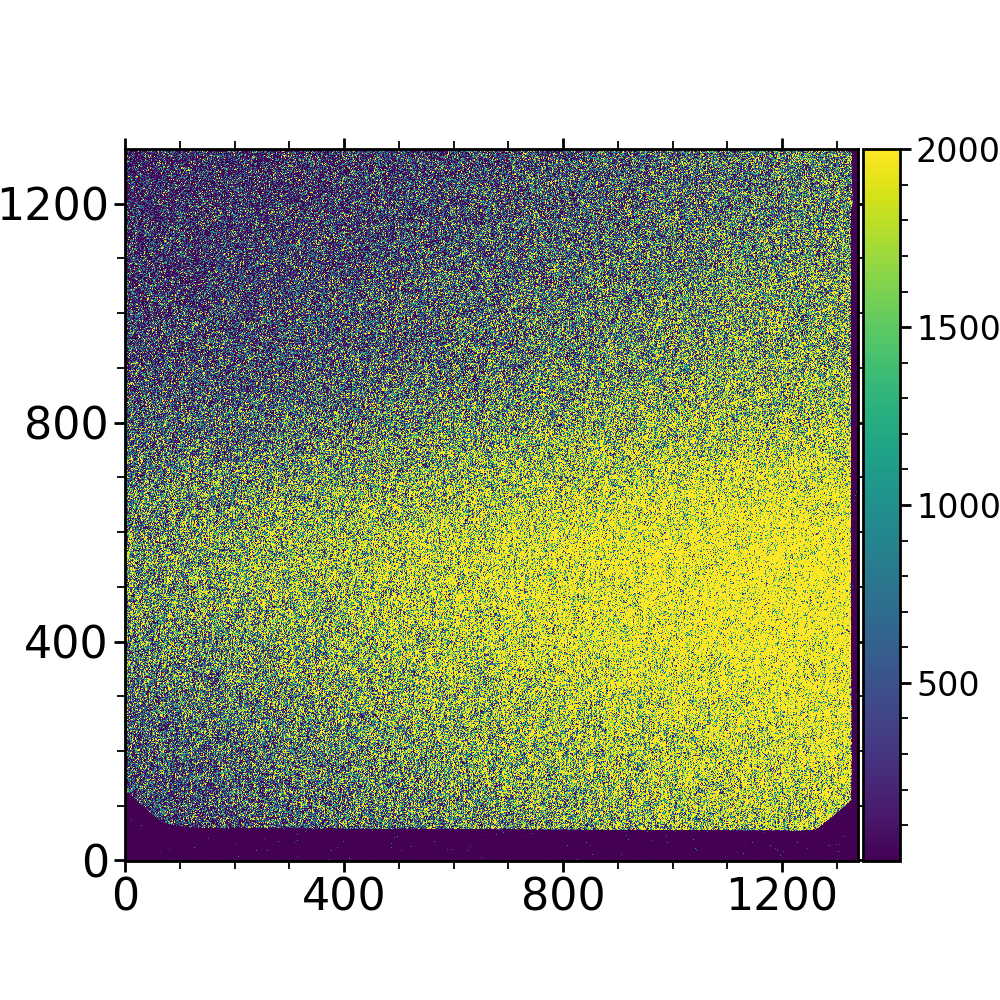}\label{subfig:fr160}}\\
    \subfloat[]{\includegraphics[width=1.685 in]{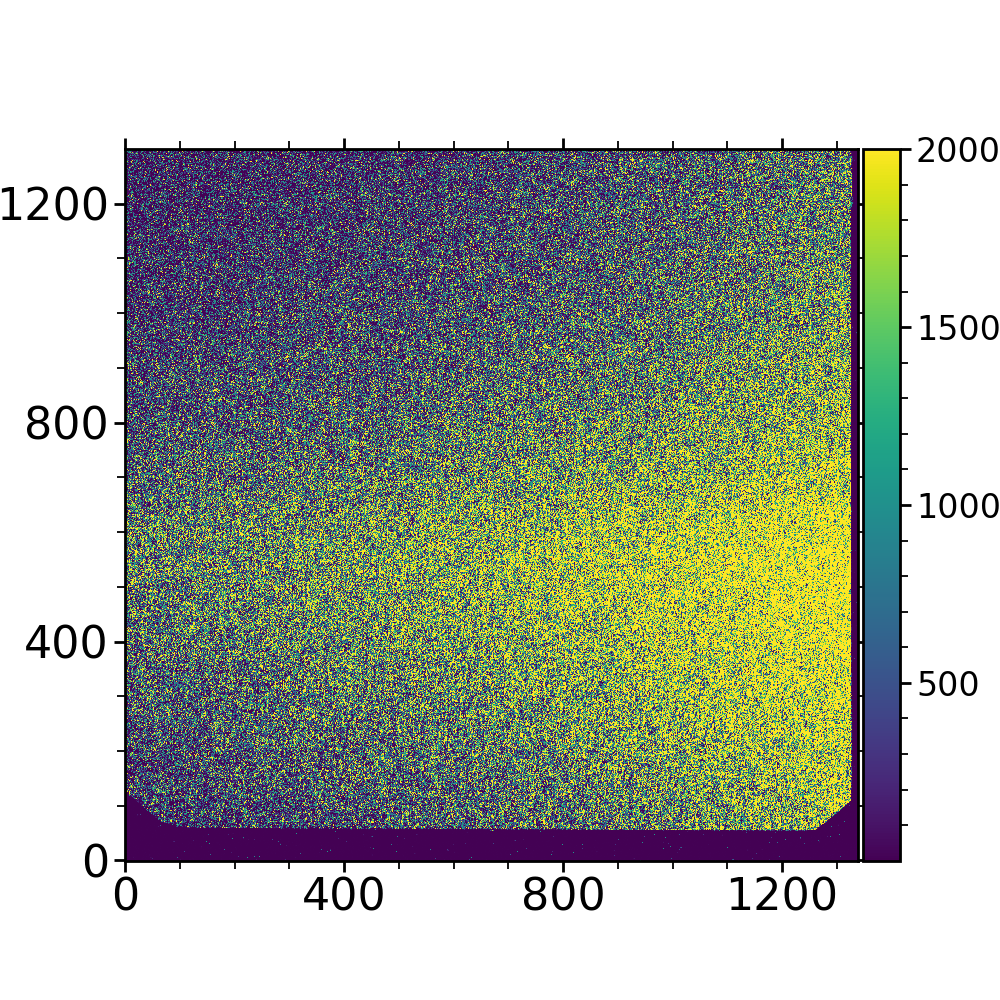}\label{subfig:fr500}}
    \caption{Intensity (arb. unit) on the detector of Sample 1 (a) before deposition; the intensity is too low to be seen on this scale; (b) 78 s after the start of deposition; and (c) 758 s after the start of deposition (kinetic steady state). The \(x\) and \(y\) axes are the pixel numbers.}
    \label{fig:3frames}
\end{figure}

\begin{figure}
    \centering
    \includegraphics[width=3.375in]{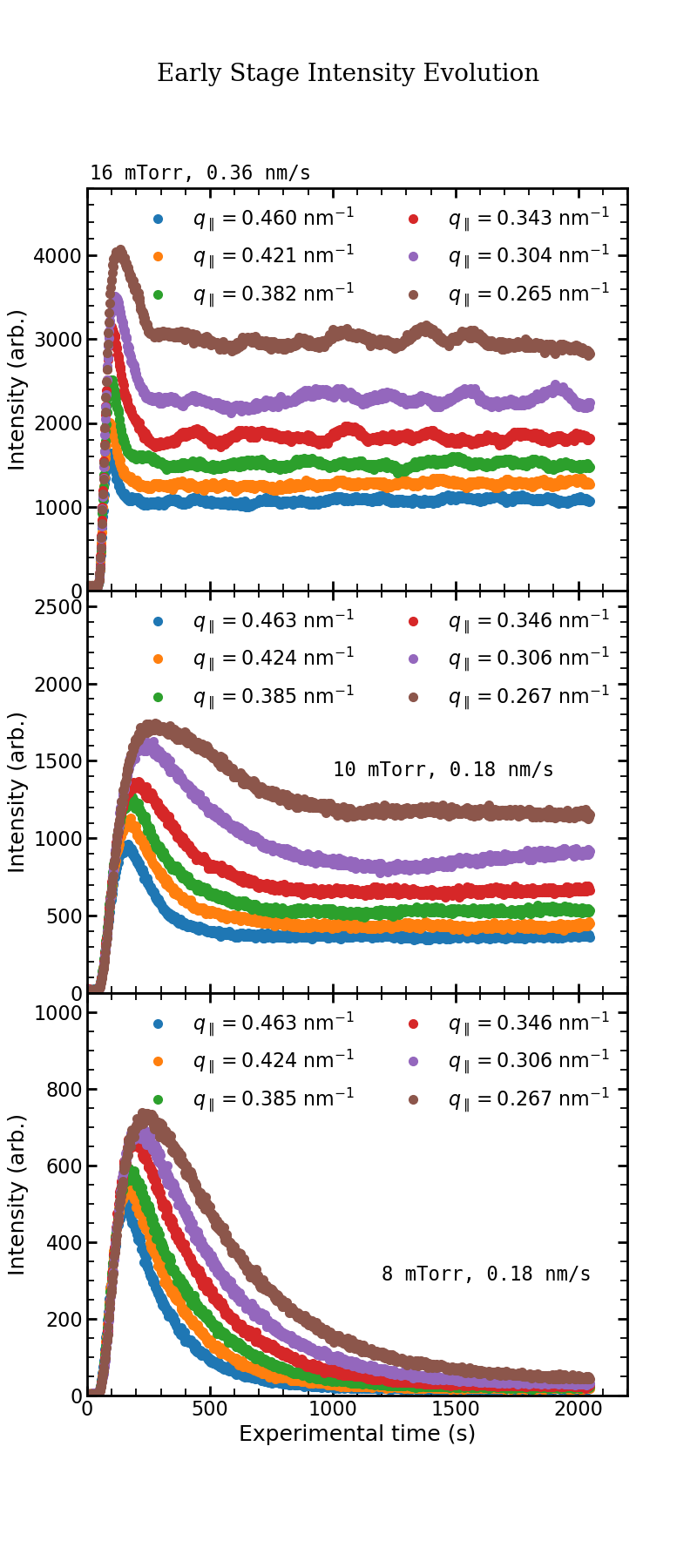}
    \caption{Speckle averaged intensity vs. time in the early stage deposition. ``Effective time'' is not used here.}
    \label{fig:I-t}
\end{figure}

Fig.\ref{fig:3frames} shows how the scattering pattern for Sample 1 changes with time. It is seen that the scattering intensity increases to a maximum (Fig.\ref{subfig:fr160}) before reaching a steady state (Fig.\ref{subfig:fr500}). In order to demonstrate the kinetics behavior, the evolution of speckle averaged intensity along the Yoneda wings as a function of time are plotted in Fig.\ref{fig:I-t}. Deposition starts at about 40 s. Each dot in Fig.\ref{fig:I-t} is calculated as the averaged intensity over a square of pixels centered at the corresponding \(\mathbf{q}\) at the Yoneda wing. Therefore, each curve in Fig.\ref{fig:I-t} represents one of the $q_{\parallel}$ positions along the Yoneda wing. In the regime of the detector,
\begin{equation}
    q_{\parallel} \approx q_y = \frac{2\pi}{\lambda}\cos(\alpha_f)\sin(\psi)
    \label{eq:q_para}
\end{equation}
because $q_x \ll q_y$. In each curve, the intensity increases rapidly in time up to a peak and then descends to a steady-state value. Moreover, the peak intensity is reached at a later time for lower \(q_{\parallel}\) values than for higher \(q_{\parallel}\)'s. When a curve with position \(q_{\parallel}\) reaches the peak, the roughness on the length scale \(\frac{2\pi}{q_{\parallel}}\) is maximum. Focusing on the kinetic steady-state behavior, we find that positions with lower \(q_{\parallel}\) stabilize later than the higher \(q_{\parallel}\) positions. This behavior is consistent with mound formation and coarsening, so that the kinetic properties reach a steady state later at longer length scales. Comparing the results among the three pressures, we see that the time to reach the kinetic steady state increases with decreasing pressure, while the final scattering intensity decreases with decreasing pressure. This is consistent with the known result that the grown surface is smoother at lower pressures \cite{zhou2010pressure}.  \par

The time to reach the peak intensity is denoted as $T_{max}$ and the time to reach the steady-state intensity (kinetic steady state) is called $T_{f}$. These two time parameters, $T_{max}$ and $T_{f}$, characterize the kinetic evolution at each wavenumber during the thin film growth. On account of the early-stage asymmetric peak and the late-stage steady state, the \(I(t)\) curves can be fit with an empirical function consisting of the sum of a lognormal distribution plus a hyperbolic tangent function
\begin{subequations}
\begin{equation}
    I(t)=I_1(t)+I_2(t)
\end{equation}
where
\begin{align}
    I_1(t)&=I_{max}\frac{1}{\sqrt{2\pi}\sigma t}e^{-\frac{(\ln(t)-\mu)^2}{2\sigma^2}}\\
    I_2(t)&=I_{\infty}\tanh(\frac{x}{T_I})
\end{align}
    \label{eq:Ifit}
\end{subequations}

\begin{figure}
    \centering
    \includegraphics[width=3.375in]{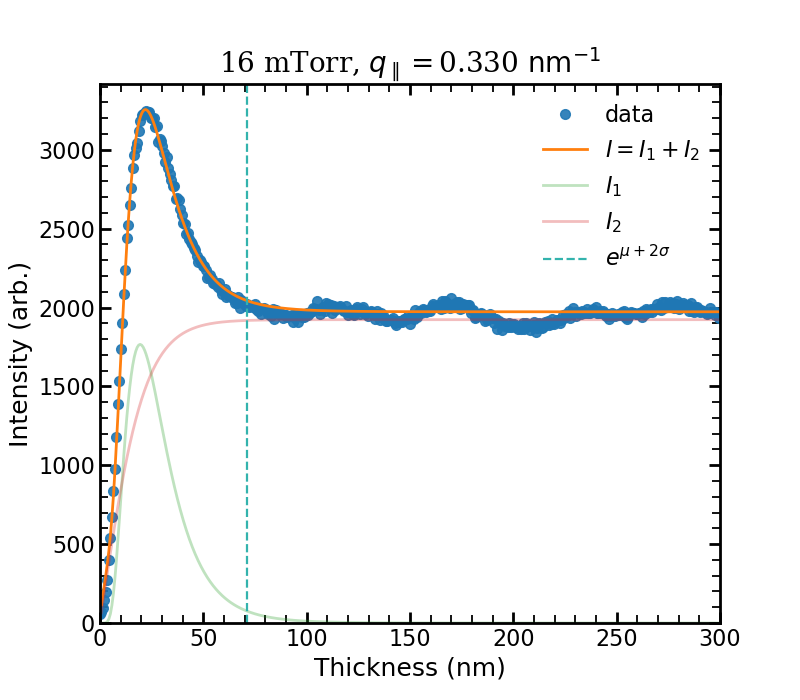}
    \caption{An example of speckle averaged intensity $I$ as a function of film thickness and curve fitting. Thickness is calculated as a product of deposition rate and deposition time.}
    \label{fig:I-t_fit}
\end{figure}

\noindent All fitting parameters in Eq.(\ref{eq:Ifit}) are $q_{\parallel}$ dependent. \(T_{max}\) is read directly as the time when the maximum intensity at a given wavenumber is reached, and \(T_{f}\) can be calculated from the curve fitting (Fig.\ref{fig:I-t_fit})

\begin{equation}
    T_{f} = e^{\mu+2\sigma}
\end{equation}

\par
\begin{figure}
    \centering
    \includegraphics[width=3.375 in]{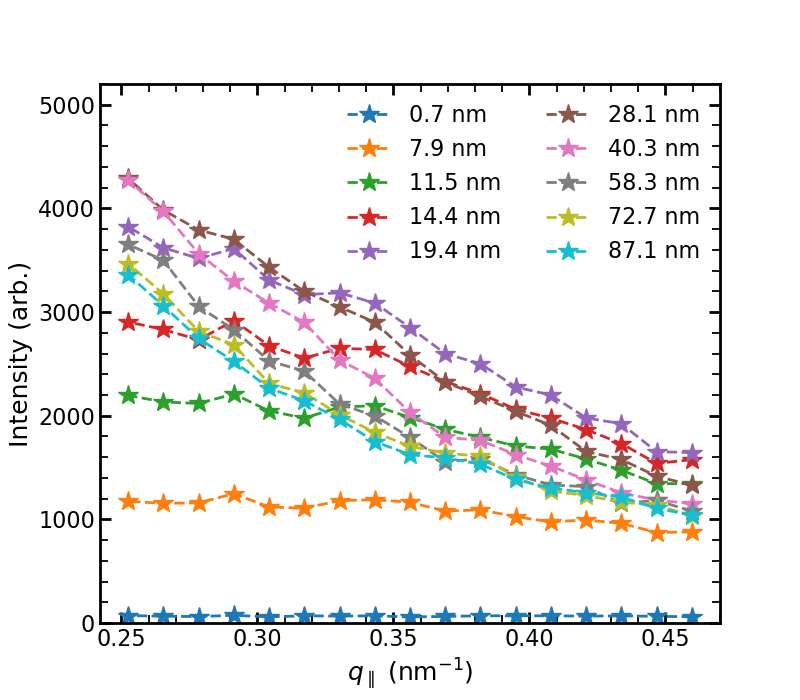}
    \caption{Speckle averaged Yoneda peak intensities of Sample 1 as a function of \(q_\parallel\) at different time (thicknesses)}
    \label{fig:I-q w/ t}
\end{figure}

On the other hand, the evolution of intensity can be replotted in a way that the $I$ vs. $q_\parallel$ curves evolve with time (Fig.\ref{fig:I-q w/ t}). At all $q_{\parallel}$ positions, the intensity increases rapidly with time at the very beginning. At low $q_{\parallel}$, the intensities rise to their peaks later than at high $q_{\parallel}$, so that crossovers between curves are observed. Ultimately the intensities reach their steady state values. The crossovers of different curves is consistent with coarsening of lateral structure during growth \cite{arslan2012evidence}. \par

\begin{figure}
    \centering
    \includegraphics[width=3.375in]{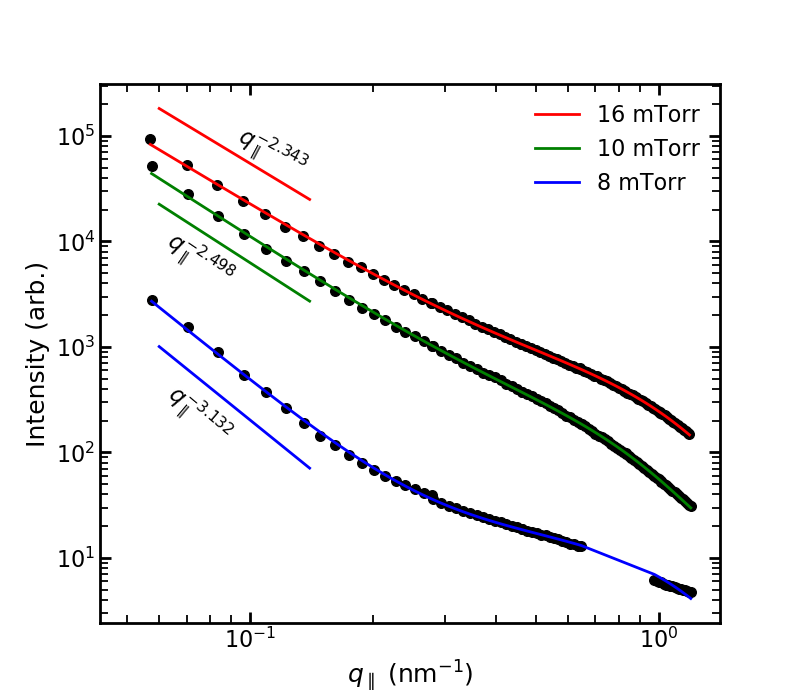}
    \caption{Speckle averaged intensity $I$ vs. $q_{\parallel}$ plots and fits in the kinetic steady state of all three samples}
    \label{fig:I-q}
\end{figure}

Scans of a broader $q_{\parallel}$ range are taken after the early stage evolution, and the results are shown in Fig.\ref{fig:I-q}. The curves are fit by a power law plus a Gaussian tail
\begin{equation}
    I(q_{\parallel}) = I_p q_{\parallel}^{-m} + I_g e^{-\frac{q_{\parallel}^2}{2\sigma^2}}
    \label{eq:power_gauss}
\end{equation}
The power law behavior in low $q_{\parallel}$ conforms to fractal growth \cite{barabasi1995fractal}, and the Gaussian tail in high $q_{\parallel}$ ranges has been observed in previous experiments \cite{rainville2015co}. As discussed in Ref.\cite{rainville2015co}, it is believed that the Gaussian component contains the bulk structure information from the near surface region. The lower intensity of the low pressure sample indicates a smoother surface at all length scales examined.

\subsection{Two-time correlation function\label{subsec:2TCF}}

The kinetic behavior only gives us the information on the average evolution. In order to go beyond the average structure, the two-time correlation function (2TCF) is applied to explore the dynamical correlation process. The 2TCF is
calculated as
\begin{equation}
    C(\mathbf{q},t_1,t_2) = \frac{\langle I(\mathbf{q},t_1)I(\mathbf{q},t_2)\rangle_{\mathbf{q}}}{\langle I(\mathbf{q},t_1)\rangle_{\mathbf{q}}\langle I(\mathbf{q},t_2)\rangle_{\mathbf{q}}}
    \label{eq:2TCF}
\end{equation}
The \(\langle I(\mathbf{q},t)\rangle_{\mathbf{q}}\) means the averaged intensity over a rectangle of pixels centered at $\mathbf{q}$. In practice, however, the Savitzky-Golay filter is applied to better estimate the normalized scattering \cite{savitzky1964smoothing}
\begin{subequations}
\begin{equation} 
    C(\mathbf{q},t_1,t_2)=\langle I'(\mathbf{q},t_1)I'(\mathbf{q},t_2)\rangle_{\mathbf{\mathbf{q}}}
\end{equation}
where
\begin{equation}
    I'(\mathbf{q},t)=\frac{I(\mathbf{q},t)}{\tilde{I}(\mathbf{q},t)}
\end{equation}
\label{eq:calc_2TCF} 
\end{subequations}
\(\tilde{I}(\mathbf{q},t)\) is composed of pixels smoothed by the 2D Savitzky-Golay filter.\par

\begin{figure}
    \centering
    \includegraphics[width=3.375in]{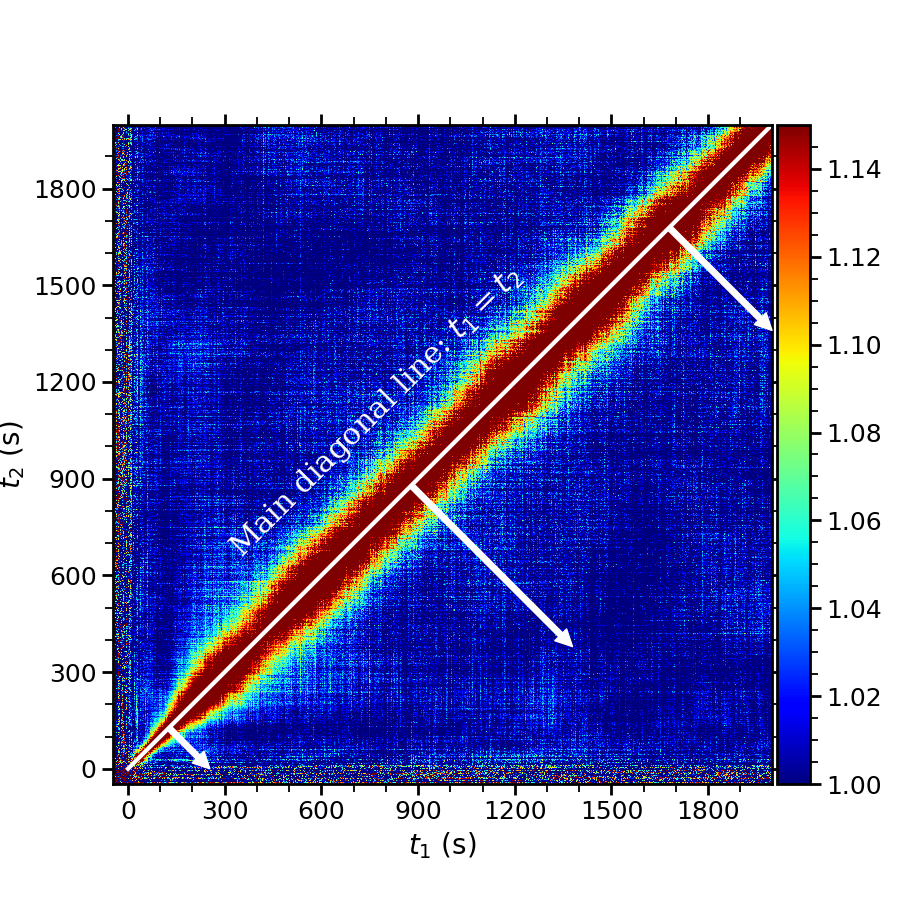}
    \caption{2TCF of Sample 2 at $q_{\parallel}=0.463~\nano\metre^{-1}$ at the Yoneda wing. \(t_1\) and \(t_2\) are deposition time. \(t_{1,2}<0\) before deposition starts. The whole figure is symmetric about the main diagonal line \(t_1=t_2\) by definition. The white arrows demonstrate the direction of slicing along the off-diagonal lines keeping \(T\) as a constant in Eq.(\ref{eq:T}).}
    \label{fig:2TCF}
\end{figure}

\begin{figure}
    \centering
    \includegraphics[width=3.375in]{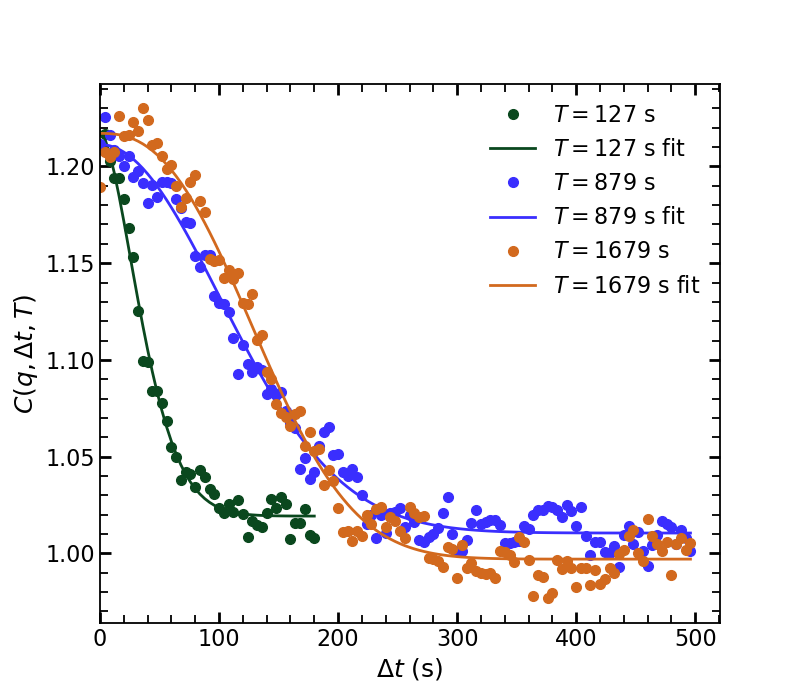}
    \caption{Off-diagonal line slices of 2TCF along the three white arrows in Fig.\ref{fig:2TCF} and the KWW fits. \(T=127~\second\) lies in the early stage, and the other two curves lie in the dynamic steady state.}
    \label{fig:comp_exp}
\end{figure}

The 2TCF of Sample 2 for $q_{\parallel}=0.463~\nano\metre^{-1}$ at the Yoneda wing is given in Fig.\ref{fig:2TCF}. The red band near the diagonal line shows the high correlation of the surface when \(t_1\) is close to \(t_2\). The `width' of the central band tells us how rapidly the surface structure changes during the growth of layers. The dynamic steady state is attained when the correlation width becomes stable, implying that the dynamical processes at the scale of \(\frac{2\pi}{q_{\parallel}}\) has reached a dynamical equilibrium. We need to go further, performing off-diagonal slices through the \(t_1=t_2\) line in order to characterize the dynamical properties. Keeping
\begin{equation}
    T=\frac{1}{2}(t_1+t_2)
    \label{eq:T}
\end{equation}
as a constant, three typical off-diagonal slices are shown in Fig.\ref{fig:comp_exp}, where \(\Delta t = \left|t_2-t_1\right|\) is the $x$-axis. The curves are fit by the Kohlrausch-Williams-Watts (KWW) form \cite{williams1970non}
\begin{equation}
    C_{\mathbf{q},T}(\Delta t) = B+\beta e^{-2(\frac{\Delta t}{\tau(\mathbf{q};T)})^n}
    \label{eq:KWW}
\end{equation}
where \(\beta\) is the contrast, \(\tau(\mathbf{q},T)\) is the correlation time, \(n\) determines the compressed ($n\geqslant 1$) or stretched ($n\leqslant 1$) exponential form. Since we fix \(q_z\) at the Yoneda wing, \(\tau(\mathbf{q},T)\) is actually \(\tau(q_\parallel,T)\) in this context. In our experiments, \(n\) is always greater than \(1\), indicating a compressed exponential relaxation. \(B\) is the baseline, which is controlled close to 1 (varied from 0.9 to 1.1 in the fittings). The compressed exponential result suggests nonlinear dynamics \cite{ulbrandt2016direct} which will be discussed further in Section \ref{sec:Sim}. 

\begin{figure}
    \centering
    \includegraphics[width=3.375in]{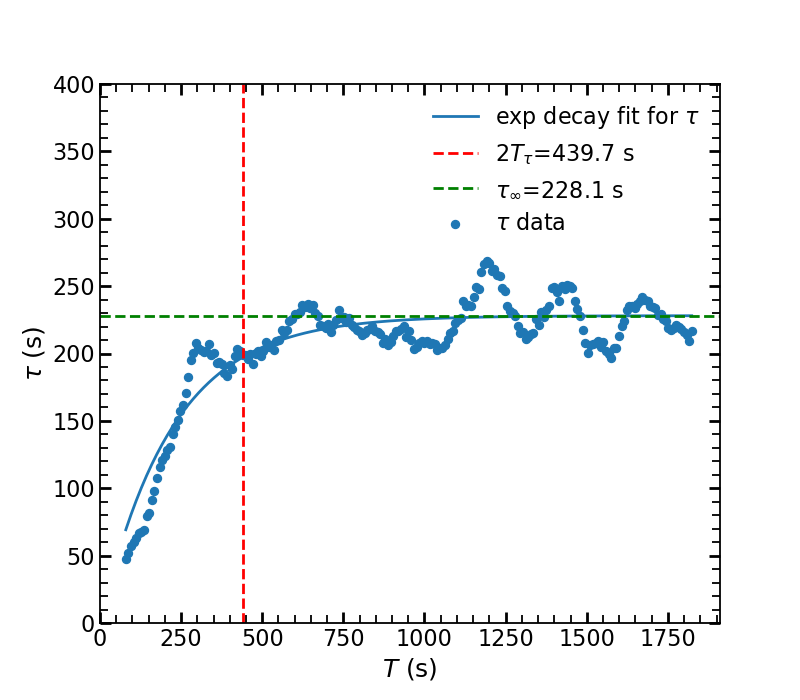}
    \caption{$\tau$-$T$ curve at $q_{\parallel}=0.346~\nano\metre^{-1}$ of Sample 2. The oscillations in the dynamic steady state are due to challenges from normalizations of the constantly evolving speckle patterns.}
    \label{fig:tau-T}
\end{figure}

Once we get all fitting parameters along the $T$ axis, the function of correlation time $\tau$ vs. $T$ is obtained (Fig.\ref{fig:tau-T}). The curves can be fit empirically using an exponential decay function:
\begin{equation}
    \tau_{q_\parallel}(T) = \tau_\infty(1-e^{-\frac{T}{T_\tau}})+\tau_0 e^{-\frac{T}{T_\tau}}
\end{equation}
The $q_\parallel$-dependent fitting parameters \(T_\tau\) and \(\tau_\infty\) are used as dynamical time parameters. The time \(2T_\tau\) approximately characterizes the time from the beginning of the deposition to reaching the dynamic steady state, and \(\tau_\infty\) characterizes the correlation time in the dynamic steady state. Note that \(T_\tau\) is the time for the surface to fundamentally change its morphology on a given length scale. Thus \(2T_\tau\) is the time it takes for the local surface morphology evolution to reach a steady state rate of change. \(\tau_\infty\) characterizes the time scale of that steady state rate of change on a given length scale.

\subsection{\label{subsec:g2}$g_2$ function}
The steady state correlation property can also be calculated according to the autocorrelation function
\begin{equation}
    g_2(\mathbf{q};\Delta t) = \langle\frac{\langle I(\mathbf{q},t)I(\mathbf{q},t+\Delta t)\rangle_t}{\langle I(\mathbf{q},t)\rangle_t^2}\rangle_\mathbf{q}
    \label{eq:g2calc}
\end{equation}
\(g_2\) analysis makes up the basis of dynamic light scattering (DLS) \cite{sutton2008review}. There are two brackets of average. $\langle\cdots\rangle_\mathbf{q}$ has the same meaning as in Eq.(\ref{eq:2TCF}). $\langle\cdots\rangle_t$ takes the average over time. Symmetrical normalization \cite{schatzel1988photon} is applied to avoid violent oscillations at large lag times, and the multi-tau method is applied for the consideration of calculation efficiency \cite{culbertson2007distributed,magatti2001fast}. The curves are fit with the KWW form (Eq.(\ref{eq:KWW})) with the baseline constrained to determine $\tau$. \par

\begin{figure}
    \centering
    \includegraphics[width=3.375in]{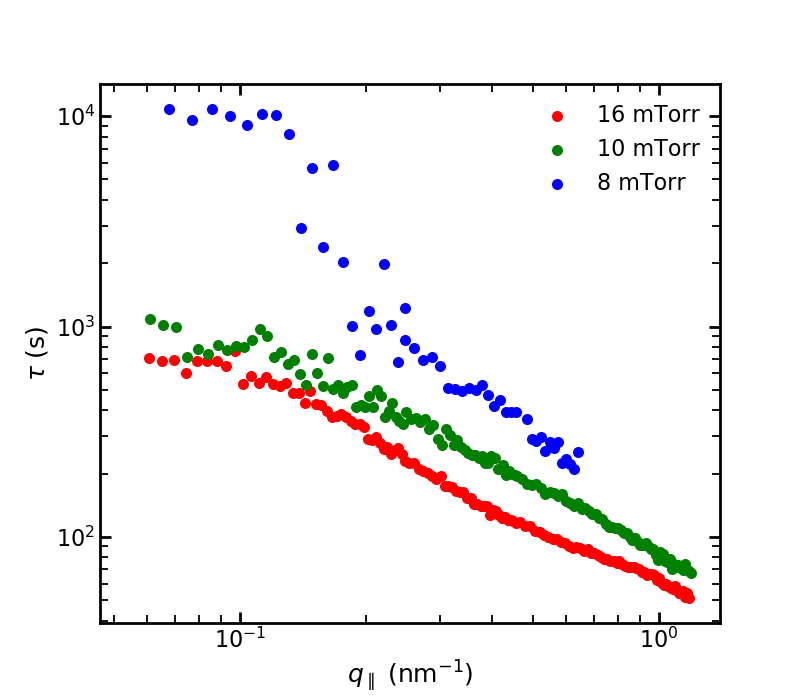}
    \caption{\(\tau\)-\(q_\parallel\) curves of three pressures given by \(g_2\) analysis}
    \label{fig:g2}
\end{figure}
The resulting \(\tau\) values are shown in Fig.\ref{fig:g2}. The high \(q_\parallel\) data of Sample 3 is missing because the intensity on the detector is too weak to calculate $g_2$ functions. \(\tau\) decreases with increasing \(q_\parallel\) and pressure, demonstrating that under lower pressure, the smoother surfaces remain correlated for a longer period of time during the growth process. The \(\tau\)'s obtained in $g_2$ functions are used as the best estimated \(\tau_\infty\) since the \(g_2\) functions are based on a large amount of steady-state data, which reflect the steady-state dynamical behavior more accurately.

\subsection{\label{subsec:result}Results and discussions}
\begin{table*}
    \centering
    \begin{tabular}{|*{4}{c|}}\hline
        Time Params & K/D & Characterization & Physical Interpretation \\\hline
        \(T_{max}(q_\parallel)\) & Kinetics & $I$-$t$ curves reach the max. & The mounds have coarsened through the length scale \(\frac{2\pi}{q_\parallel}\). \\\hline
        \(T_{f}(q_\parallel)\) & Kinetics & $I$-$t$ curves reach the steady state. & Local morphology reaches steady state at the scale. \\\hline
        \(T_{\tau}(q_\parallel)\) & Dynamics & 2TCFs reach the steady state. & Local dynamics reaches steady state at the scale. \\\hline
         & & & Surface structure at the corresponding length scale \\
        \(\tau_{\infty}(q_\parallel)\) & Dynamics & The width of 2TCFs & remains correlated over this time (thickness). \\\hline
    \end{tabular}
    \caption{Illustration of four time parameters}
    \label{tab:time_params}
\end{table*}

\begin{figure*}
    \centering
    \includegraphics[width=7in]{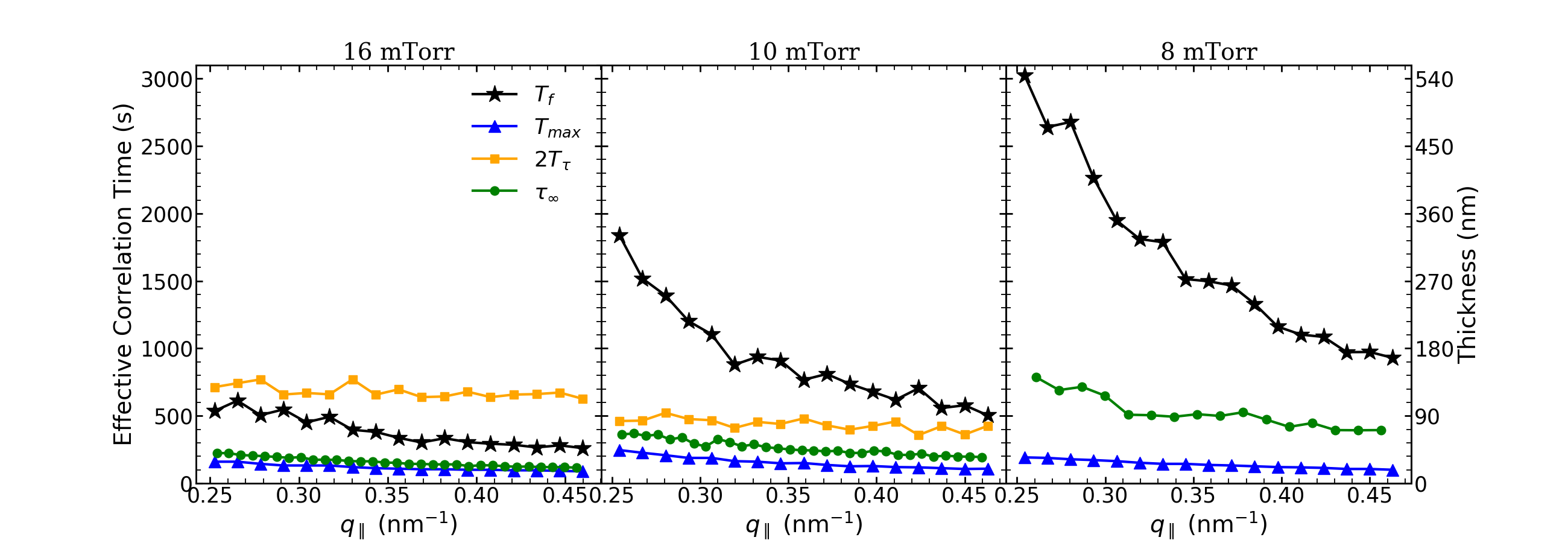}
    \caption{The \(q_\parallel\) dependence of the four experimentally determined time parameters for the 3 samples}
    \label{fig:time_params}
\end{figure*}

The overall goal of the X-ray data analysis has been to extract two kinetic parameters, \(T_{max}\) and \(T_f\), and two dynamic parameters, \(T_\tau\) and \(\tau_\infty\). A summary is given in Table \ref{tab:time_params} and the final results are given in Fig.\ref{fig:time_params}. The \(T_\tau\) for Sample 3 is absent because the surface is too smooth to extract signals with meaningful contrast in the 2TCF. Effective deposition time as mentioned in Section \ref{subsec:expdet} has been applied to balance the different deposition rates. \par

In order to demonstrate that the kinetic energy and the velocity direction of adatoms are strongly affected by pressure, let us estimate the mean free paths through which silicon and tungsten molecules travel in argon atmosphere. The mean free path of a gas molecule is calculated as:
\begin{equation}
    \bar{\lambda} = \frac{k_B T}{\sqrt{2}\sigma p}
    \label{eq:free_path}
\end{equation}
where \(k_B\) is the Boltzmann constant, \(T\) is the gas temperature, \(\sigma\) is the scattering cross section, and \(p\) is the gas pressure. The cross section can be calculated as \(\sigma=\pi d^2\), where \(d\) is the distance at which there starts to be interatomic forces between two atoms. We now estimate the scattering cross section of a tungsten or silicon atom scattered by an argon atom. The radius of one silicon atom is \(r_{\mathrm{Si}}=0.132~\nano\metre\), the radius of one tungsten atom is \(r_{\mathrm{W}}=0.141~\nano\metre\), and the radius of one argon atom is \(r_{\mathrm{Ar}}=0.071~\nano\metre\) \cite{princetonedu}. With the atomic radii, we have
\begin{subequations}
\begin{align}
    d_{\mathrm{W}} &= r_{\mathrm{W}}+r_{\mathrm{Ar}} = 0.212~\nano\metre\\
    d_{\mathrm{Si}} &= r_{\mathrm{Si}}+r_{\mathrm{Ar}} = 0.203~\nano\metre
\end{align}
\end{subequations}
The experiment was done under room temperature so \(T=298~\kelvin\). Substituting the experimental pressures in Eq.(\ref{eq:free_path}), we have the mean free paths of Si and W atoms in Table \ref{tab:mfp}.

\begin{table}
\begin{tabular}{|*{4}{c|}}\hline
    \(p\) & 16 mTorr & 10 mTorr & 8 mTorr \\ \hline
     & & & \\
    \(\bar{\lambda}_\mathrm{W}\)(cm) & 0.966 & 1.55 & 1.93 \\ \hline
     & & & \\
    \(\bar{\lambda}_\mathrm{Si}\)(cm) & 1.05 & 1.69 & 2.11 \\ \hline
\end{tabular}
\caption{Mean free paths of W and Si atoms at three different pressures}
\label{tab:mfp}
\end{table}

Since the mean free path of a molecule is much less than the working distance \(10~\centi\metre\), the adatoms will experience multiple collisions on their way to the surface. It has been hypothesized that atom collisions in the vapor play an increasingly important role with increasing pressure in the experimental pressure regime \cite{zhou2010pressure}. Furthermore, the mean free path varies from about 1 cm at 16 mTorr to about 2 cm at 8 mTorr. Thus increased pressure can potentially lead to more random deposition direction and to nanoclusters forming in the vapor. Collisions also reduce the kinetic energy of the adatoms, presumably leading to smaller diffusion lengths. Less diffusion and greater cluster sizes could then result in a rougher surface with a morphology which stabilizes more quickly. We note that simulations and calculations have shown the importance of adatom kinetic energy to diffusion lengths on the surface \cite{morales2014hot,gao2012transient}. However, unlike previous works which examined the effect of this behavior on cluster formation, here we evoke kinetic energy-enhanced diffusion as a mechanism to smoothen the amorphous growth surface. Another effect of increased adatom scattering in the vapor above the surface is to randomize their arrival directions. Atom steering due to surface interactions is a potentially important mechanism of surface roughening \cite{van1999steering,shevchik1973growth}. It is more effective for lower energy adatoms and grazing incidence approach to the surface, so that it would increase with increasing pressure. \par

Comparing the results among three pressure conditions, the following observations can be made:
\begin{asparaenum}
\item At 16 mTorr, it is seen that \(T_{f}<2T_\tau\). Physically, this means that the kinetic steady state is reached earlier than the fluctuation dynamics reaches a steady state. At the measured length scales, this phenomenon physically means that dynamical processes driven by adatom absorption and diffusion continue to evolve after the average morphology is no longer changing during the thin film growth. Zhou \textit{et al.} \cite{zhou2010pressure} have suggested that the deposited particles are in clusters at high pressures. Also, according to the mean free path calculation above, particles at higher pressures are deposited with lower kinetic energy. Under such circumstances, we hypothesize that the effects of adatom diffusion on the surface structure are weaker, causing the local morphology to become saturated earlier than in the case at lower pressures. While deposited nanoclusters are still contributing to evolution of the dynamic processes, the surface has already reached a steady state kinetic roughness morphology.
\item \(q_\parallel\) dependence. Apart from \(T_\tau\), the other three parameters show obvious \(q_\parallel\) dependence, and \(T_f\) shows the strongest \(q_\parallel\) dependence. Therefore, the time scale to reach the kinetic steady state varies significantly with lateral length scale. From the strong \(q_\parallel\) dependence of \(T_f\) and weak \(q_\parallel\) dependence of \(T_\tau\), we could deduce that, if this behavior continues beyond the observed \(q_\parallel\) window, there will be a crossing point at \(q_\parallel<0.23~\nano\metre^{-1}\) in the $16$ mTorr experiment and at \(q_\parallel>0.47~\nano\metre^{-1}\) in the $10$ mTorr experiment, at which length scale the kinetic and dynamic steady states are reached simultaneously. The \(q_\parallel\) independence of \(T_\tau\) suggests that the fluctuation dynamics reaches the steady state at the same time on a wide range of length scales.
\item Pressure dependence. Pressure-dependent behavior is most strongly seen in \(\tau_\infty\) and \(T_{f}\). The time scales decrease with increasing pressure. As the surface becomes rougher and the kinetic energy of adatoms becomes lower with increasing pressure, the steady-state local morphology will form faster. Under low pressure conditions, a smooth surface with sufficient dynamical relaxation preserves the existing morphology over a greater time of continuing film deposition.
\end{asparaenum}

\section{\label{sec:Sim}Theory and Simulation}

In this section we introduce a continuum model for comparison with the kinetic and dynamic properties discussed in Section \ref{sec:Anal}. $x$ and $y$ axes are equivalent in the experimental sputter deposition geometry, so an isotropic equation should be applied. The Kuramoto-Sivashinsky (KS) equation was invented to describe spontaneous instabilities in nonequilibrium physical and chemical systems, such as chemical turbulence and burning fronts \cite{kuramoto2003chemical,sivashinsky1983instabilities}. Besides isotropy of \(\mathbf{x}\), the translational symmetry of \(t\) and \(h\), and inverse symmetry of \(t\) should also be taken into consideration. All physically possible terms up to the fourth order are included in the following differential equation \cite{raible2001amorphous}:
\begin{equation}
\begin{split}
    & \frac{\partial H}{\partial t} = a_1\nabla^2H+a_2\nabla^4H \\
    & +a_3\nabla^2(\nabla H)^2+a_4(\nabla H)^2+a_5M+F+\eta
\end{split}
    \label{eq:raible_full}
\end{equation}
with 
\begin{equation}
M = \begin{vmatrix}
\partial_x^2H & \partial_y\partial_xH \\
\partial_x\partial_yH & \partial_y^2H
\end{vmatrix}
\end{equation}
In Eq.(\ref{eq:raible_full}), \(H\) is the actual height of the surface, \(F\) is the average growth velocity and noise \(\eta\) conforms to stochastic uniform distribution. \par

The term with \(M\) can be neglected compared to the \(\nabla^2H\) term \cite{raible2001amorphous}. By defining \(h=H-Ft\) as the height measured from the average surface level, the equation we adopted for simulation is expressed as
\begin{equation}
    \frac{\partial h}{\partial t} = a_1\nabla^2h+a_2\nabla^4h+a_3\nabla^2(\nabla h)^2+a_4(\nabla h)^2+\eta
    \label{eq:sim}
\end{equation}
where the random noise \(\eta\) satisfies
\begin{subequations}
\begin{align}
    \langle\eta(\mathbf{x},t)\rangle &= 0 \\
    \langle\eta(\mathbf{x},t)\eta(\mathbf{x'},t')\rangle &= 2D\delta^d(\mathbf{x}-\mathbf{x'})\delta(t-t')
\end{align}
\end{subequations}
We used a square grid of size \(L\times L\) with \(L=1024\), and set the time step to be \(\Delta t = 0.01\). One length unit is set as \(1\) nm and one time unit represents \(1\) s.\par

Physical backgrounds of the five terms are discussed in Ref.\cite{barabasi1995fractal,raible2001amorphous,raible2000amorphous_theory,raible2000amorphous_minimal,lai1991kinetic,pelliccione2008evolution}. The signs of coefficients in Eq.(\ref{eq:sim}) are
\begin{equation*}
    a_1,a_2,a_3<0\text{ and }a_4>0
\end{equation*}
The coefficient \(a_1<0\) because incident particles on their way to the surface will be deflected toward mounds by atomic attractions \cite{shevchik1973growth}. Thus, more particles arrive at the surface where \(\nabla^2h<0\) (Fig.2 in Ref.\cite{raible2001amorphous}), contrary to the effect of a positive Laplacian term in the Edwards-Wilkinson (EW) equation. In the EW equation, a positive \(\nabla^2\) term comes from surface desorption, but this mechanism is highly suppressed in the present case because the experiment is done under room temperature. The \(\nabla^4h\) term results from the surface diffusion in Mullin's equation \cite{mullins1957theory}. The third term \(a_3\nabla^2(\nabla h)^2\) arises from the equilibration of the inhomogeneous surface concentration, and \(a_4(\nabla h)^2\) can be derived from the variation of surface inclinations. \((\nabla h)^2\) is usually referred to as the Kardar-Parisi-Zhang (KPZ) term \cite{kardar1986dynamic} and \(\nabla^2(\nabla h)^2\) is referred to as the conserved KPZ term. The KPZ term is non-conserved so it can lead to excess velocity. \par

\begin{figure}
    \centering
    \includegraphics[width=3.375in]{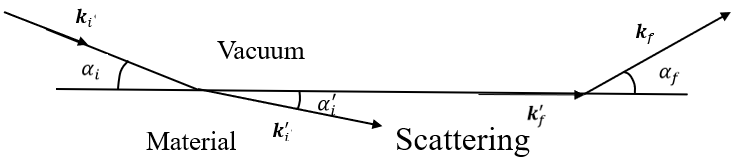}
    \caption{Schematic diagram of scattering inside the material}
    \label{fig:qz'}
\end{figure}

\begin{figure}
    \centering
    \includegraphics[width=3.375in]{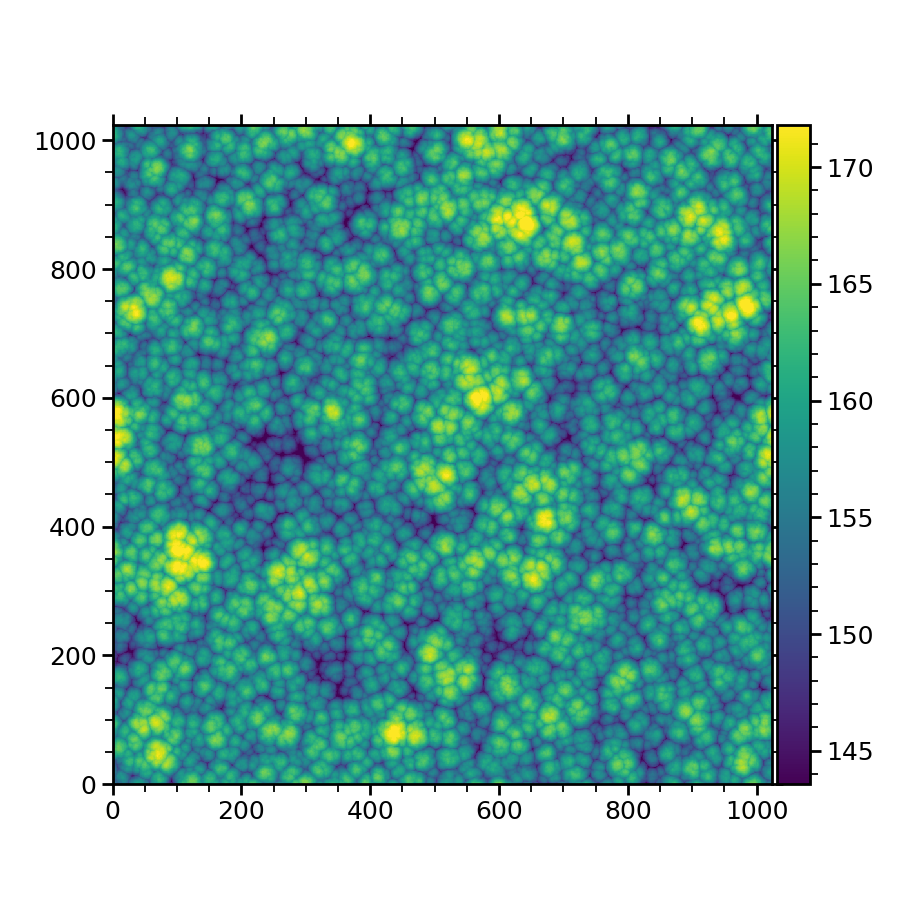}
    \caption{Simulation result \(h(x,y)\) in the real space at \(t=2000\second\). The unit of height is nm.}
    \label{fig:realsp_sim}
\end{figure}

We investigated a wide range of parameters for Eq.(\ref{eq:sim}). According to Ref.\cite{raible2000amorphous_theory}, \(a_1=-Fb\), where \(b\) is the difference between the typical range of interatomic forces and the equilibrium distance of the deposited particles. Even if taking the ten times of the atomic radius to be \(b\), that is to say, \(b=10r_{\mathrm{W}}=1.41~\nano\metre\), \(a_1\) will only reach a value of \(-0.25~\nano\metre^2\cdot\second^{-1}\), whose absolute value is too small to generate a peak in the early stage \(I\)-\(t\) curves even if the other parameters are also small in magnitude. In order to resemble the \(I\)-\(t\) behavior in Fig.\ref{fig:I-t}, \(\left|a_1\right|\) cannot be too small because a negative \(a_1\) generates the initial instability. There will not be a peak in early stage \(I\)-\(t\) curves unless a sufficiently negative \(a_1\) is applied to compete with the terms \(a_2\nabla^4h\) and \(a_4(\nabla h)^2\) that stabilize the morphology evolution and cause the scattering intensity to saturate. Correspondingly, although some theories were developed to predict the other parameters \cite{raible2001amorphous}, the core features in the experiment must be reproduced with a proper selection of the other coefficients compatible with \(a_1\). Ultimately, we found the best general agreement with experiment using the following parameters:
\begin{equation}
\begin{split}
    a_1=-0.64~\nano\metre^2\cdot\second^{-1}&\text{;~~~}
    a_2=-1~\nano\metre^4\cdot\second^{-1}\text{;}\\
    a_3=-2~\nano\metre^4\cdot\second^{-1}&\text{;~~~}
    a_4=0.1~\nano\metre\cdot\second^{-1}\text{;~and~}\\
    D=3.8\times10^{-3}&~\nano\metre^4\cdot\second^{-1}
\end{split}
\label{eq:coeffs_sim}
\end{equation}
Real space result of $t=2000~\second$ is shown in Fig.\ref{fig:realsp_sim}. Mounds are found on top of longer wavelength roughness, and these mounds coalesce and split during the growth.

After simulating the surface in real space, a transformation to coherent X-ray scattering data is required. The scattering pattern is therefore obtained from the simulated surface using the formula \cite{sinha1988x,mokhtarzadeh2017simulations}
\begin{equation}
\begin{split}
    I(q_x,q_y)&\propto\frac{1}{q_{z}'^{2}L^{2}}\left| \int\int\,\mathrm{d}x\,\mathrm{d}y e^{-iq_{z}'h(x,y)}e^{-i(q_xx+q_yy)}\right|^2\\
     &=\frac{1}{q_{z}'^{2}L^{2}}\left|\mathcal{F}\{e^{-iq_z'h(x,y)}\}\right|^2
\end{split}
\end{equation}
where \(q_z'\) is the \(z\) component of the wave vector change \(\mathbf{q'}\) inside the material (Fig.\ref{fig:qz'}):
\begin{equation}
    \mathbf{q'} = \mathbf{k_f'}-\mathbf{k_i'}
\end{equation}
\(\mathbf{k}_i\) and \(\mathbf{k}_f\) are incident and scattered beams outside the material. \(\mathbf{k}_i'\) is refracted beam entering the material and \(\mathbf{k}_f'\) is the scattered beam inside the material. For scattering at the Yoneda wing, \(\mathbf{k}_f'\) propagates along (parallel to) the surface. According to Snell's law and scattering geometry,
\begin{equation}
\begin{cases}
    n = \cos \alpha_c \\
    n\cos \alpha_i' = \cos \alpha_i \\
    q_z' = k_i'\sin \alpha_i' \\
\end{cases}
\end{equation}
With small-angle approximation, \(\alpha_i'\) can be expressed as
\begin{equation}
    \alpha_i'^2 = \alpha_i^2-\alpha_c^2
\end{equation}
Therefore,
\begin{equation}
    q_z' \approx \frac{2\pi\alpha_i'}{\lambda} = \frac{2\pi\sqrt{\alpha_i^2-\alpha_c^2}}{\lambda}
    \label{eq:qz'}
\end{equation}
Substituting \(\alpha_i=0.45\degree\), \(\alpha_c=0.40\degree\) and \(\lambda=\frac{hc}{E}\) where \(E=7.38~\kilo\electronvolt\) in Eq.(\ref{eq:qz'}), we have \(q_z'=0.134~\nano\metre^{-1}\).

\begin{figure}
    \centering
    \includegraphics[width=3.375in]{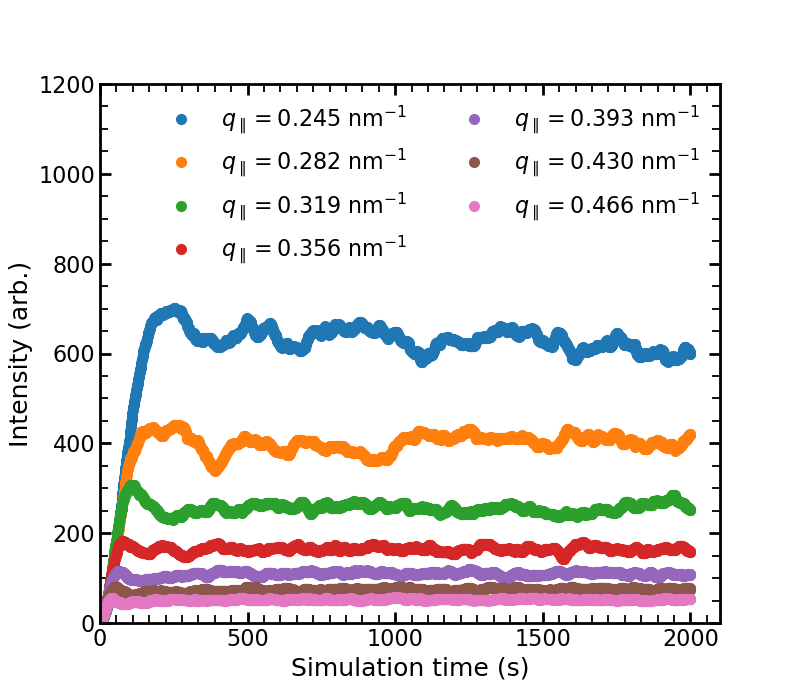}
    \caption{Simulation result: speckle averaged intensity \(I\)-\(t\) in the early stage}
    \label{fig:I_t_sim}
\end{figure}

\begin{figure}
    \centering
    \includegraphics[width=3.375in]{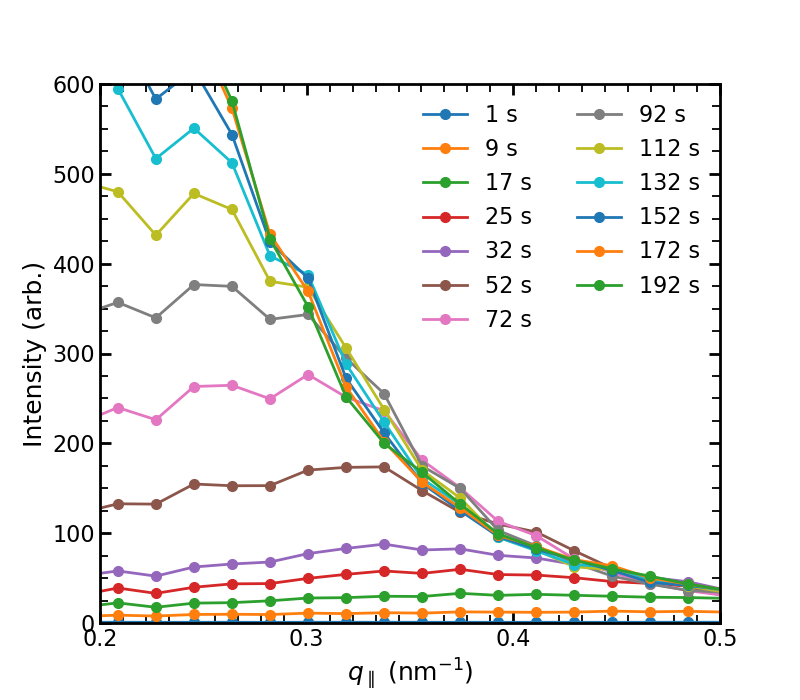}
    \caption{Simulation result: speckle averaged intensity \(I\)-\(q_{\parallel}\) in the early stage evolving with time}
    \label{fig:I_q_early_sim}
\end{figure}

\begin{figure}
    \centering
    \includegraphics[width=3.375in]{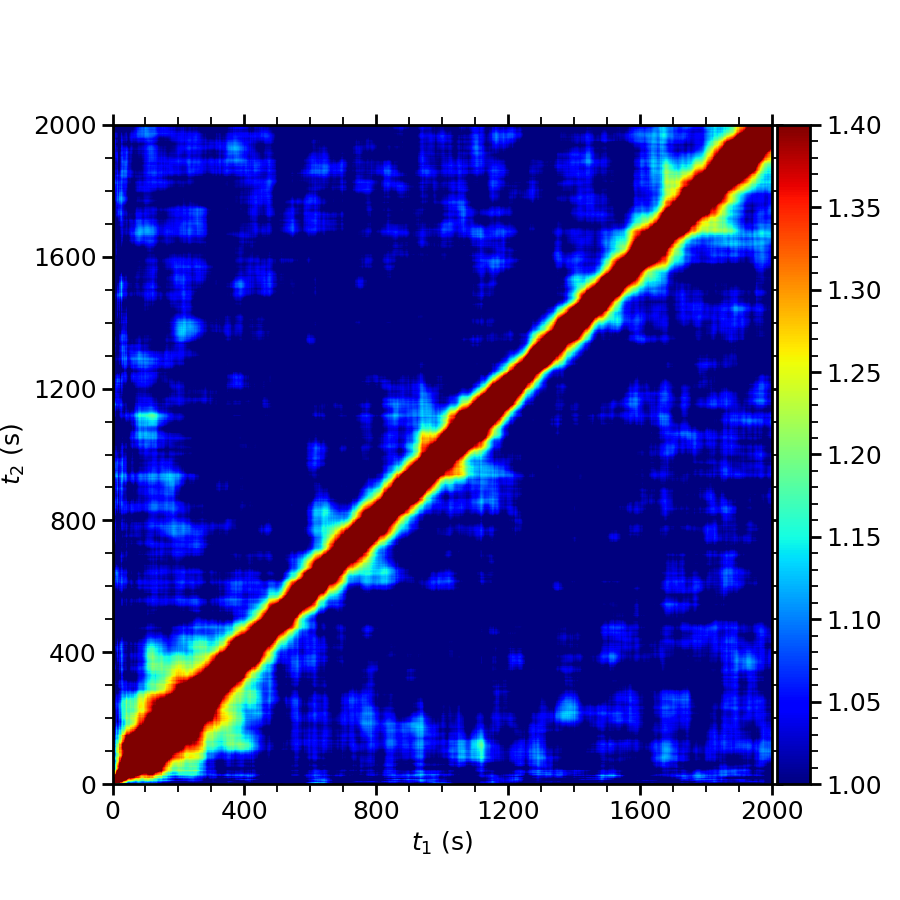}
    \caption{2TCF from simulation at \(q_\parallel=0.264~\nano\metre^{-1}\)}
    \label{fig:2TCF_sim}
\end{figure}

\begin{figure}
    \centering
    \includegraphics[width=3.375in]{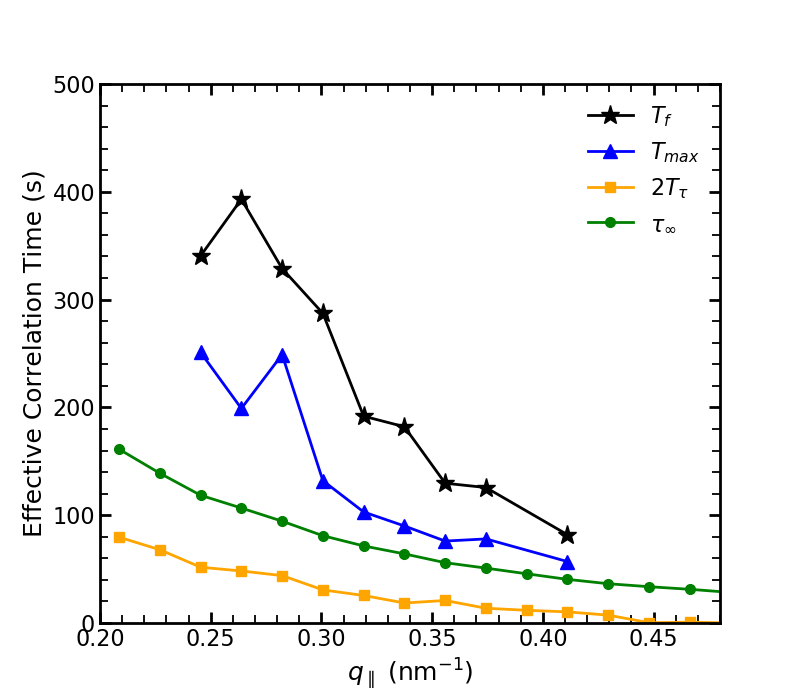}
    \caption{The \(q_\parallel\) dependence of the four time parameters from the simulation}
    \label{fig:time_params_sim}
\end{figure}

Now we are able to characterize the kinetic properties from the calculated intensity evolution. The early stage kinetic behavior is shown in Fig.\ref{fig:I_t_sim} and Fig.\ref{fig:I_q_early_sim}. To compare the dynamics between the simulation and the experiment, speckle correlation analysis was performed for the simulation in the same way as for the X-ray data. Fig.\ref{fig:2TCF_sim} gives an example of what the 2TCF looks like, and the final results of the four time parameters are plotted in Fig.\ref{fig:time_params_sim}.\par

The simulation is successful to some extent because many features have resemblances to what was observed in the experiment. In Fig.\ref{fig:I_t_sim}, it is firstly observed in the continuum model that a peak develops in the structure factor at early times. The peak in \(I(q_\parallel)\) (Fig.\ref{fig:I_q_early_sim}) shifts to smaller \(q_\parallel\) with time and there is overlapping of subsequent curves. The \(q_\parallel\) dependence of \(T_f\) and \(\tau_\infty\) resembles Fig.\ref{fig:time_params}.\par

However, the simulation results also display a few disagreements with the experiment. For the kinetics, peaks are observed in \(I(t)\) in only part of the \(q_\parallel\) range instead of the entire \(q_\parallel\) range from $0.23$ to $0.47~\nano\metre^{-1}$, and the peaks in Fig.\ref{fig:I_q_early_sim} are seen at greater \(q_\parallel\)'s compared with the experimental result. Focusing on the dynamic behavior, \(2T_\tau\) decreases so significantly with \(q_\parallel\) that the surface reaches the dynamic steady state very quickly after the deposition starts. Therefore \(T_f\) is always greater than \(2T_\tau\). Furthermore, the strong \(q_\parallel\) dependence of \(T_{max}\) and \(T_\tau\) does not accord with the experiments.

\section{\label{sec:Conc}Conclusion}

In conclusion, the Co-GISAXS technique is used to study the kinetic and dynamic behavior during a-WSi\(_2\) thin film deposition as a function of growth pressure. After analyzing height-height structure factors and correlation functions, we characterize the kinetic and dynamic processes by extracting four time parameters, \(T_f\), \(T_{max}\), \(T_\tau\) and \(\tau_\infty\) with physical interpretations in Table \ref{tab:time_params}. The peaks in the \(I\)-\(t\) curves lead to \(T_{max}\), implying that at each length scale, the roughness reaches a maximum before relaxing down to a steady state. Kinetic properties reach a steady state later at longer length scales because of mound formation and coarsening. Comparing the kinetic behavior at different pressures, we can conclude that the surface is rougher at higher pressure, as is expected because of decreasing kinetic energy and increasingly random directions of adatoms. This phenomenon can also be explained by the hypothesis that clusters of greater sizes aggregate in the vapor at higher pressures, so the deposited particles are actually nanoclusters instead of single atoms \cite{zhou2010pressure}. With such an assumption, the mean free path would be even smaller. The surface at 8 mTorr is too smooth to calculate the autocorrelation function. Therefore we could not extract \(T_\tau\) value for 8 mTorr. \par

We then discuss the \(q_\parallel\) and pressure dependence in Fig.\ref{fig:time_params}. Apart from \(T_\tau\), the other three parameters show obvious \(q_\parallel\) dependence under each pressure condition, and \(T_f\) shows the strongest \(q_\parallel\) dependence. \(T_f\) varies significantly with \(q_\parallel\) because the average structure at greater lateral sizes of course takes longer time to saturate, but \(T_\tau\) does not show such strong \(q_\parallel\) dependence, presumably because dynamical processes driven by atomic arrival, diffusion and incorporation do not rely on the length scale we detect. \(T_{max}\) increases with increasing length scale because it takes longer to form larger mounds, and the surface structure at longer length scales stays correlated longer than at shorter length scales. \par 

As for pressure dependence, a marked finding is the inversion between \(T_f\) and \(2T_\tau\). At pressures as high as 16 mTorr, the kinetic steady state is found to be reached even earlier than the fluctuation dynamics reaches a steady state. At high pressures, the adatoms are deposited with lower kinetic energy, so they diffuse over a short length scale before they stop, and it will take a shorter time for the local morphology to reach a steady state. However, if local dynamical processes depend on surface roughness, then they will take longer to stabilize under higher pressures because of the higher roughness that develops. At low pressures, the surface roughness is small, so the dynamical processes do not change much from what they were on the original flat surface. Coherent X-ray scattering essentially reflects the height fluctuation, so \(2T_\tau\) decreases a little bit from 16 to 10 mTorr, and \(2T_\tau>T_f\) at 16 mTorr. \(T_{max}\) hardly changes with pressure, which shows that early stage coarsening through a definite lateral length scale does not depend on pressure. A smoother surface structure remains correlated for a longer time, so \(\tau_\infty\) increases with decreasing pressure. \par

In the late stages of the experiment (Fig.\ref{fig:I-q}), the power law behavior of \(I(q_\parallel)\) indicates a self-affine structure on the measured length scales, which may be described by a continuum model. Moreover, the non-exponential intensity evolution and the compressed exponential decay in Fig.\ref{fig:comp_exp} require nonlinear terms. Therefore, Eq.(\ref{eq:sim}) was adopted to describe the early stage kinetic and dynamic behavior. The set of parameter in Eq.(\ref{eq:coeffs_sim}) represents the best overall behavior among all of our simulations. This simulation result resembles many features in the experiment. In part of the \(q_\parallel\) ranges, \(I(t)\) reaches a peak before reaching a lower steady-state value. The peak in \(I(q_\parallel)\) shifts to smaller \(q_\parallel\) with time and the subsequent curves overlap, although the peaks are seen at greater \(q_\parallel\)'s compared with the experiment. The \(q_\parallel\) dependence of \(T_f\) and \(\tau_\infty\) is consistent with the experiment. However, the strong \(q_\parallel\) dependence of \(T_{max}\) is not consistent with the experiment. The least agreement with experiment comes from \(T_\tau\), which almost vanishes in high \(q_\parallel\). In addition, attempts to simulate the pressure dependence behavior by simply varying the simulation coefficients were not successful. \par

Improvements to the continuum model could be possible. According to Ref.\cite{raible2000amorphous_theory}, \(a_5=Fb^2\). If we adopt the `nanoclusters forming in vapor' hypothesis, \(b\) will be in the order of 1 nm or above, so \(a_5M\) in Eq.(\ref{eq:raible_full}) cannot be neglected. The nonuniformity of cluster sizes at high pressures may require a more sophisticated treatment of the noise term. Besides, higher order derivatives may also contribute to kinetics and dynamics, and different magnitudes of discretization could be applied to different pressure conditions if we could confirm the hypothetical dependence of the mean cluster sizes of deposition on pressures. \par

The nature of the continuum model has itself limited the range of simulated dynamic behavior and surface evolution. Eq.(\ref{eq:sim}) has contained all possible terms to \(\mathcal{O}(\nabla^4,(\nabla h)^2)\) for an isotropic growth process with definite physical sense. The shortcomings noted above were observed for all simulation parameter sets examined. Restricted by local behavior, local continuum models cannot describe nonlocal mechanisms like shadowing. Furthermore, continuum models fail to track short length-scale processes of particles or clusters because of discretization, losing part of the correlation behavior especially at higher \(q_\parallel\)'s. In other words, local morphology below a nanometer is lost because of the discretization. An alternative solution could be atomistic models, which might overcome some of these limitations.

\begin{acknowledgments}
We thank Peco Myint for use of his XPCS analysis program and for helpful conversations. The component of this work at BU was partly supported by the National Science Foundation (NSF) under Grant No. DMR-1709380. At UVM, J.G.U. and R.L.H. were partly supported by the U.S. Department of Energy (DOE) Office of Science under Grant No. DE-SC0017802. This research used resources of the Advanced Photon Source, a U.S. Department of Energy (DOE) Office of Science User Facility operated for the DOE Office of Science by Argonne National Laboratory under Contract No. DE-AC02-06CH11357.
\end{acknowledgments}

\bibliography{Early_stage_paper.bib}

\end{document}